\documentclass[preprint, prX]{revtex4}
\usepackage{mathrsfs,amsfonts}
\usepackage{bm}
\usepackage{amsmath}

\begin{document}
\title 
{Plasmonic metamaterial enhanced axionic magnetoelectric effect}
\author{Yong Zeng$^{1}$ and Hou-Tong Chen$^{2}$}
\address{$^1$ Theoretical Division, MS B213, Los Alamos National Laboratory, Los Alamos, NM 87545, USA\\
$^2$ MPA-CINT, MS K771, Los Alamos National Laboratory, Los Alamos, NM 87545, USA}
\input epsf
\begin{abstract}
Axionic electrodynamics predicts many peculiar magnetoelectric-based properties. Hitherto, simple structures such as one-dimensional multilayers were employed to explore these axionic magnetoelectric responses, and Fabry-P\'{e}rot interference mechanism was frequently applied to augment these effects. In this Letter, we propose a new mechanism, metamaterial-enhanced axionic magnetoelectric response, by taking advantage of intense enhancement of localized electromagnetic
fields associated with plasmonic resonances. Through numerical simulations, we show that plasmonic metamaterial can enhance axionic magnetoelectric effect by two orders of magnitude.
\end{abstract}
\maketitle

Axion is a pseudoparticle postulated by the Peccei-Quinn theory in 1977 to resolve the strong CP (CP standing for Charge Parity) problem in quantum chromodynamics \cite{Peccei}. To describe its interaction with electromagnetic (EM) field, the ordinary Maxwell Lagrangian of classical electromagnetism should be modified by including an axionic term proportional to $\theta\mathbf{E}\cdot\mathbf{B}$ \cite{Wilczek}. This so called axionic electrodynamics predicted many new and novel physics mainly because the additional term gives rise to magneto-electric effects. For example, electric charges induce magnetic monopoles and vice versa in the presence of a planar domain wall across which $\theta$ jumps \cite{Sikivie,Huang}. For recent developments in axionic electrodynamics, please refer to Ref \cite{Itin} and the references given therein.

Recently, axionic electrodynamics found its physical reality in condensed matter physics. It is suggested that one can use axionic electrodynamics to describe EM properties of low-energy topological insulators \cite{qi1}. A topological insulator is a material that behaves as an insulator in its interior but contains conducting states near its surface. This topological current sheet leads to interesting boundary conditions \cite{Obukhov}. As a result, topological insulators possess quantized magnetoelectric effects, which result in exotic phenomena \cite{Karch,qi2}: It is predicted that an electric charge near a topological surface state can induce an image magnetic monopole charge \cite{qi3}; Three-dimensional topological insulators may present repulsive Casimir forces \cite{Grushin1,Grushin2}; Considerable magneto-optical Kerr effects and Faraday effects of thin-film topological insulators are predicted theoretically and demonstrated experimentally \cite{Tse1,Tse2,Maciejko,Jenkins}.

Most current studies with regard to axionic electrodynamics are limited to simply structures such as stratified multilayers, and usually take advantage of the Fabry-P\'{e}rot mechanism to achieve enhanced axionic responses \cite{Tse2,Maciejko}. On the other hand, plasmonic metamaterials have been applied to strongly manipulate matter-wave interactions in the past decade \cite{john,solymar,Schuller}. In this Letter, we propose to apply plasmonic metamaterials to enhance magnetoelectric effects of axionic media. For the specific design presented below, the magnetoelectric response is found to be increased by two orders of magnitude.

Assuming a time dependence of $e^{-i\omega t}$, the axionic electrodynamics is described by the standard Maxwell's equation \cite{Wilczek}
\begin{eqnarray}
&&\nabla\times\mathbf{E}=i\omega\mathbf{B},\:\:\:\:\:\nabla\cdot\mathbf{D}=0, \cr &&\nabla\times\mathbf{H}=-i\omega\mathbf{D},\:\:\:\nabla\cdot\mathbf{B}=0,
\label{eq1}
\end{eqnarray}
together with an unique constitutive relation
\begin{equation}
\mathbf{D}(\omega)=\epsilon_{0}\epsilon\mathbf{E}-\beta\eta_{0}\mathbf{B},\:\:\mathbf{H}(\omega)=\frac{\mathbf{B}}{\mu_{0}\mu}+\beta\eta_{0}\mathbf{E},
\label{eq2}\end{equation}
where the two $\beta$ terms stand for the axion fields, and $\eta_{0}=\sqrt{\epsilon_{0}/\mu_{0}}$ is the admittance of free space. A material with above constitutive relation will be referred to as an axionic medium. Strictly speaking, axionic media are also bi-anisotropic media whose constitutive relations can have up to 36 variable moduli \cite{Obukhov,Kong}. For a plane wave solution in a homogenous axionic medium, one can prove that
\begin{equation}
\mathbf{E}=-\frac{\omega}{k}\mathbf{\hat{k}}\times\mathbf{B},\:\:\:\mathbf{H}=\frac{\omega}{k}\mathbf{\hat{k}}\times\mathbf{D},
\label{eq3}\end{equation}
where $\mathbf{k}$ being the wave vector. Consequently, the electric field $\mathbf{E}$ is orthogonal to the magnetic induction $\mathbf{B}$, but does not parallel the electric displacement $\mathbf{D}$.

\begin{figure}[b]
\epsfxsize=250pt \epsfbox{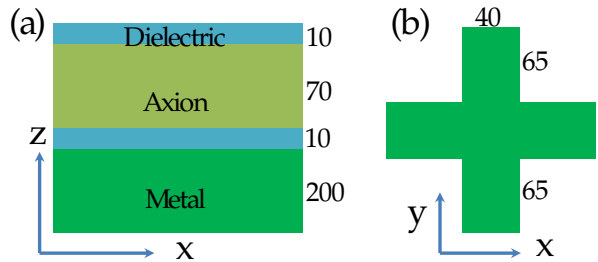} \vspace*{-7.5cm}
\caption{Schematics of (a) Structure A and (b) the plasmonic metamaterial used in Structure B. The metamaterial is a square array of metallic crosses, has a thickness of 100 nm and a lattice constant of 2 $\mu$m. All dimensions are in nanometers.}
\end{figure}

Without loss of generality, let us consider a non-magnetic (with $\mu=1$) axionic structure. Its vector wave equation is given by
\begin{equation}
\nabla\times\nabla\times\mathbf{E}-\frac{\omega^{2}}{c^{2}}\epsilon(\mathbf{r},\omega)\mathbf{E}=i\omega\mu_{0}\left[\mathbf{J}^{i}+\eta_{0}\mathbf{E}\times\nabla\beta(\mathbf{r})\right],
\label{eq4}
\end{equation}
where $\mathbf{J}^{i}$ represents the current source in infinity which generates the incident wave. To solve this equation, one can image an auxiliary system where the axionic medium is replaced by a normal dielectric with identical permittivity, and further define a Green's function $\underline{\mathbf{G}}^{d}$ which satisfies
\begin{equation}
\nabla\times\nabla\times\underline{\mathbf{G}}^{d}(\mathbf{r},\mathbf{r}')-\frac{\omega^{2}}{c^{2}}\epsilon(\mathbf{r},\omega)\underline{\mathbf{G}}^{d}(\mathbf{r},\mathbf{r}')=\underline{\mathbf{I}}\delta(\mathbf{r}-\mathbf{r}').
\label{eq5}
\end{equation}
Consequently, the solution of Eq. (\ref{eq4}) can be written as
\begin{equation}
\mathbf{E}(\mathbf{r})=i\omega\mu_{0}\int \underline{\mathbf{G}}^{d}(\mathbf{r},\mathbf{r}')\cdot\mathbf{J}^{i}d\mathbf{r}'+i\omega\mu_{0}\eta_{0}\int \underline{\mathbf{G}}^{d}(\mathbf{r},\mathbf{r}')\cdot\mathbf{E}\times\nabla'\beta(\mathbf{r}')d\mathbf{r}'.
\label{eq6}
\end{equation}
The first term on the right hand side describes a process in which an incident wave is scattered by the auxiliary structure and does not contribute to the magnetoelectric effect. Using the fact that $\nabla\beta(\mathbf{r})$ is nonzero only at interfaces across which $\beta$ jumps, one can reformulate the second term as
\begin{equation}
i\omega\mu_{0}\eta_{0}\sum_{k}\delta\beta_{k}\int_{s_{k}} \underline{\mathbf{G}}^{d}(\mathbf{r},\mathbf{r}')\cdot\left[d\mathbf{s}_{k}'\times\mathbf{E}(\mathbf{r}')\right].
\label{eq7}
\end{equation}
where $s_{k}$ stands for the $k$-th interface across which $\beta$ changes. Since $d\mathbf{s}_{k}'\times\mathbf{E}(\mathbf{r}')$ can modify the electric field polarization, this term therefore is the only source for the axionic magnetoelectric effects. More importantly, one may significantly enhance these effects by increasing the electric fields at the axinoic interfaces.

To demonstrate this new mechanism, we consider two axionic structures. Structure A, shown in Fig.1(a), consists of four homogeneous dielectric/axion/dielectric/metal slabs. Structure B is identical to Structure A except it has an additional plasmonic metamaterial on top. The metamaterial is a square array of metallic crosses and is shown schematically in Fig.1(b). For both structures, the incident plane wave is assumed to be $x$-polarized and propagates along the $-z$ direction. Since the plasmonic metamaterial possesses a four-fold rotational symmetry along the $z$ axis, it alone does not alter the wave polarization. To quantitatively measure the axionic magnetoelectric response, one may use the $y$-polarized reflection coefficient $R_{y}(\omega)$
\begin{equation}
R_{y}(\omega)=\left|\frac{E_{y}^{r}(\omega)}{E_{x}^{i}(\omega)}\right|^{2},
\label{eq8}
\end{equation}
since no wave can transmit through the metallic ground layer. Here $E_{y}^{r}$ is the $y$-polarized component of the reflected electric field, and $E_{x}^{i}$ is the incident electric field.

To numerically simulate these two structures, we develop a three-dimensional finite-difference time-domain (FDTD) algorithm (Supplementary Material) \cite{code}. In the simulations, the dielectric has a constant permittivity of 2.28 (corresponds to Al$_{2}$O$_{3}$). Additionally, the permittivity of the metal is described by a Drude model, $1-\omega_{p}^{2}/(\omega^{2}+i\omega\gamma_{m})$. For comparison purpose (discussed below), the bulk plasma frequency $\omega_{p}$ is chose to be $1.37\times10^{4}$ THz and the decay rate $\gamma_{m}=0.41$ THz. In order to suppress the staircase error of FDTD and achieve guaranteed accuracy, the axionic medium is assumed to have an identical permittivity as the dielectric and a tunable $\beta$.

\begin{figure}[t]
\epsfxsize=300pt \epsfbox{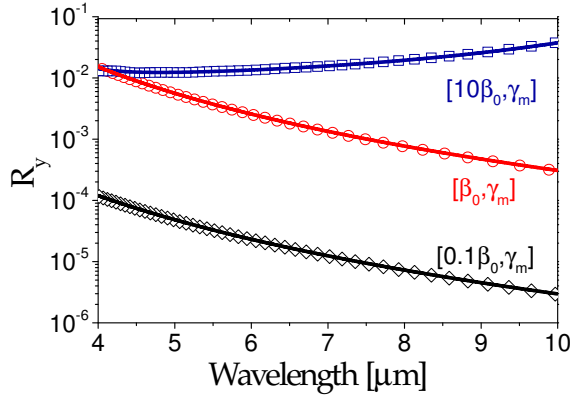} \vspace*{-8.0cm}
\caption{The axionic magnetoelectric responses of Structure A. Two different approaches are used: The FDTD algorithm (scatters) and an analytical transfer-matrix approach (solid lines). The two quantities inside the bracket represent the $\beta$ value of the axionic medium and the decay rate of the metal respectively. Here $\beta_{0}=2.32$ and $\gamma_{m}=0.41$ THz.}
\end{figure}

Fig. 2 shows the axionic magnetoelectric response of Structure A with different $\beta$. For such a simple multilayer structure, an analytical transfer matrix method can be employed to compute $R_{y}$ \cite{Tse1}. Alternatively, one may use the FDTD algorithm. The numerical result, plotted with scatters, is in excellent agreement with its analytical counterpart. In the wavelength range of interest, $R_{y}$ is found to be small and vary monotonously.

Using the FDTD algorithm, we calculate $R_{y}$ of Structure B and the results are depicted in Fig.3. Clearly, the plasmonic metamaterial has a profound influence on the axionic magnetoelectric response, and much bigger $R_{y}$ are obtained in Structure B for two different $\beta$. Consistent with the theory above, the strongly localized evanescent fields around the metallic metamaterial enhance the magnetoelectric effect significantly, and the strongest response appears around the plasmonic resonant wavelength. For example, $R_{y}$ has a maximal value of 0.44 at 6.13 $\mu$m wavelength for the axionic medium with a $\beta$ of 3.23. The corresponding Structure A, in sharp contrast, possesses a tiny $R_{y}$ of $0.002$ at the same wavelength. Therefore, the localized surface plasmonic resonance enhances the axionic magnetoelectric response by about 200 times. Additionally, the plasmonic resonant wavelength is found to depend on the $\beta$ value of the axionic medium. When $\beta$ decreases from 3.23 to 0.323, the resonant wavelength blue shifts from 6.13 $\mu$m to 5.8 $\mu$m.

\begin{figure}[t]
\epsfxsize=300pt \epsfbox{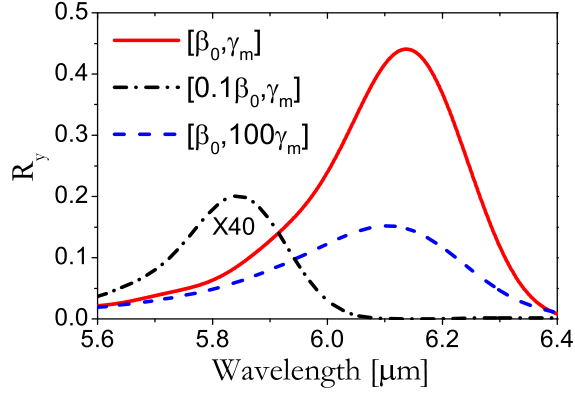} \vspace*{-8.0cm}
\caption{Enhanced magnetoelectric responses from Structure B, with different combinations of $\beta$ value and metallic decay rate.}
\end{figure}

The metal used in the above simulations is a hypothetical material with a negligible absorption loss. It is well known that metamaterial may absorb EM energy strongly around its plasmonic resonance. For example, a metamaterial perfect absorber can be designed to convert EM energy to heat very efficiently \cite{landy,chen,yong}. To study the effect of metallic losses on the axionic magnetoelectric response of Structure B, we use genuine gold to replace the ideal metal. The gold permittivity shares an identical Drude model as the ideal one except that its decay rate is now $100\gamma_{m}$. The numerical result is plotted with dashed curve in Fig.3. As expected, the magnetoelectric response is degraded due to the metallic absorption and $R_{y}$ is decreased from 0.44 to 0.15 at the resonant wavelength. On the other hand, because of the strong near-field concentration around the metamaterial, Structure B still presents a profounder magnetoelectric response than Structure A.

To qualitatively study the role of plasmonic metamaterial in the enhanced magnetoelectric effect, one may use a discrete dipole approximation by treating each metallic cross as an electric dipole with polarization $\mathbf{p}$ and polarizability $\alpha$ (which is scalar in the $xy$ plane because the metallic cross is four-fold symmetric) \cite{bohren}. The total reflected wave, under a normal incidence $\mathbf{E}^{i}e^{ik_{0}z}$, is then given by (Supplementary Material)
\begin{equation}
e^{-ik_{0}z}\left\{\underline{\mathbf{R}}(k_{0})\cdot\mathbf{E}^{i}-\frac{i\omega}{2\eta_{0}A_{cell}}\left[\underline{\mathbf{R}}(k_{0})+\underline{\mathbf{I}}\right]\cdot\mathbf{p}\right\}.
\label{eq9}
\end{equation}
Here $A_{cell}$ is the unit cell area of the cross array, $k_{0}$ is the free-space wavenumber, and $\underline{\mathbf{R}}(k_{0})$ is the reflection tensor of the multilayer substrate under a normal incidence. As suggested by Fig.2, the off-diagonal components of $\underline{\mathbf{R}}(k_{0})$ are quite small. Consequently $\mathbf{p}$ should have a considerable cross-polarized component so that the cross-polarized reflected field can be significant. Furthermore, as implied by the energy conservation law, the field corresponding to the co-polarized component of $\mathbf{p}$ should destructively interference with $\underline{\mathbf{R}}(k_{0})\cdot\mathbf{E}^{i}$ so that a significant amount of energy can be transferred to the cross polarization \cite{chen,yong}.

The polarization $\mathbf{p}$ possessed by each metallic cross can be solved self-consistently and is given by (Supplementary Material)
\begin{equation}
\mathbf{p}=\zeta\left[\underline{\mathbf{I}}-\zeta\underline{\mathbf{R}}^{t}\right]^{-1}\cdot[\underline{\mathbf{I}}+\underline{\mathbf{R}}(k_{0})]\cdot\mathbf{E}^{i},
\label{eq10}
\end{equation}
where $\zeta$ contains the localized surface plasmonic response, and the evanescent wave contributions are absorbed by $\underline{\mathbf{R}}^{t}$. As a result, the resonant frequencies of the whole structure are determined by $\underline{\mathbf{I}}-\zeta\underline{\mathbf{R}}^{t}=0$. Furthermore, one can recast $\left[\underline{\mathbf{I}}-\zeta\underline{\mathbf{R}}^{t}\right]^{-1}$ as $\underline{\mathbf{I}}+\sum(\zeta\underline{\mathbf{R}}^{t})^{n}$, and interprets it as multiple reflections between the dipole array and the substrate. All in all, the metamaterial-enhanced magnetoelectric process can be qualitatively described as: The initial polarization of the metallic cross induced by the incident field is rotated and amplified through each reflection between the metallic metamaterial and the axionic substrate. The final $\mathbf{p}$ therefore does not parallel the incident polarization and contains a considerable cross-polarized component.

To further boost the axionic magnetoelectric response, one may optimize the plasmonic metamaterial design by using low-loss metals or more suitable geometries. For example, one can bring close the axionic medium and the metamaterial so that electric field at the axionic interfaces will be stronger because of the local field concentration. At a distance of 2 nm, we numerically find that $R_{y}$ can be bigger than 0.9, implying that more than 90\% incident energy has been rotated to the $y$ polarization. It is likely that perfect polarization conversion can be achieved by using this mechanism \cite{rogacheva,hao,mutlu,Grady}.

To sum up, plasmonic metamaterials are proposed to enhance axionic magnetoelectric effects. Through numerical simulations, it is found that more than two order of enhancement can be achieved by exciting localized plasmonic resonance. Our results suggest that axionic metamaterials may be used to design magnetic metamaterials, chiral metamaterials, and electromagnetic devices such as isolators or ultrathin waveplates. To experimentally examine our propose, one may use Cr$_{2}$O$_{3}$ or Fe$_{2}$TeO$_{6}$ which carries an axionic piece in its paramagnetic phase \cite{Raab}, or topological insulator such as Bi$_{1-x}$Sb$_{x}$ alloy \cite{Hsieh} or Bi$_{2}$Se$_{3}$ crystal \cite{Jenkins} coated with a thin magnetic film.

We thank Dr. Jinjie Liu of the Delaware State University for his invaluable discussions. We acknowledge support from the LANL LDRD program. This work was carried out under the auspices of the National Nuclear Security Administration of the U.S. Department of Energy at Los Alamos National Laboratory under contract No. DE-AC52-06NA25396.

\newpage
\numberwithin{equation}{section}
\section{Supplementary material I: FDTD method}
Below we will develop a general finite-difference time-domain (FDTD) numerical algorithm \cite{Taflove} to simulate an arbitrary three-dimensional axionic structure.

Let us consider a system consists of two axionic media with different $\beta$. One can rewrite the Amp\`{e}re's law as
\begin{equation}
-i\omega\epsilon_{0}\epsilon\mathbf{E}=\nabla\times\left(\frac{\mathbf{B}}{\mu_{0}\mu}\right)+\eta_{0}(\beta_{+}-\beta_{-})\delta(n)\mathbf{e}_{n}\times\mathbf{E},
\label{eqs1}
\end{equation}
where $\epsilon$ and $\mu$ are position dependent, $\mathbf{e}_{n}$ being the normal direction of the interface pointing from $\beta_{-}$ medium to $\beta_{+}$ medium. It is important to mention that the second term on the right-hand side is nonzero only at the interface, and is the exclusive origin of the axionic magnetoelectric effect since the tangential electric field will be rotated by $90^\circ$. The equation above also implies that an axionic medium behaviors as an ordinary dielectric in its interior, and its axionic properties only appears at an interface across which $\beta$ jumps.

Without loss of generality, we assume both media are non-magnetic and $\mu=1$. By introducing a new field $\mathbf{F}=\mathbf{B}/\mu_{0}$, one can reformulate the above equation as
\begin{equation}
\nabla\times\mathbf{F}=-i\omega\epsilon_{0}\epsilon\mathbf{E}-\eta_{0}(\beta_{+}-\beta_{-})\delta(n)\mathbf{e}_{n}\times\mathbf{E}.
\label{eqs2}\end{equation}
Together with the Faraday's law
\begin{equation}
\nabla\times\mathbf{E}=i\omega\mu_{0}\mathbf{F},
\label{eqs3}\end{equation}
we can derive the remaining two Maxwell's equations, $\nabla\cdot\mathbf{F}=0$ and $\nabla\cdot\mathbf{D}=0$. Eqs (\ref{eqs2},\ref{eqs3}) therefore are sufficient to describe the axionic electrodynamics. In a non-magnetic ordinary dielectric, one can prove that $\mathbf{F}$ is equivalent to the magnetic field $\mathbf{H}$. To simplify our discussion further, the axion values $\beta$ are assumed to be real and independent of the EM frequency. The permittivity $\epsilon(\omega)$, on the other hand, can be complex and frequency dependent. To handle a complex $\epsilon(\omega)$, one can employ the auxiliary differential equation approach \cite{Taflove} by rewriting
\begin{equation}
-i\omega\epsilon_{0}\epsilon(\omega)\mathbf{E}=-i\omega\epsilon_{0}\epsilon_{e}\mathbf{E}+\mathbf{J},
\label{eqs4}\end{equation}
where $\epsilon_{e}$ is real and constant, and $\mathbf{J}$ can be interpreted as the polarization current.

Following the standard FDTD technique and transforming the continuous space-time to a discrete space-time, one can discretize Eq. (\ref{eqs2}) as
\begin{equation}
(\epsilon_{0}^{2}\overline{\epsilon}_{e}^{2}+\eta^{2})\left(\begin{array}{c}E_{s}\\E_{p}\end{array}\right)^{n+1}=\mathcal{A}\left(\begin{array}{c}(\overline{\nabla\times\mathbf{F}})_{s}-\overline{J}_{s}\\(\overline{\nabla\times\mathbf{F}})_{p}-\overline{J}_{p}\end{array}\right)^{n+1.5}
+\mathcal{B}\left(\begin{array}{c}E_{s}\\E_{p}\end{array}\right)^{n}
\label{eqs5}
\end{equation}
with
\begin{equation}
\eta=\frac{(\beta_{+}-\beta_{-})\eta_{0}\delta t}{2\delta l},\:\:\mathcal{A}=\delta t\:\left(\begin{array}{cc}\epsilon_{0}\overline{\epsilon}_{e}&-\eta\\\eta&\epsilon_{0}\overline{\epsilon}_{e}\end{array}\right),
\mathcal{B}=\left(\begin{array}{cc}\epsilon_{0}^{2}\overline{\epsilon}_{e}^{2}-\eta^{2}&-2\eta\epsilon_{0}\overline{\epsilon}_{e}\\2\eta\epsilon_{0}\overline{\epsilon}_{e}&\epsilon_{0}^{2}\overline{\epsilon}_{e}^{2}-\eta^{2}\end{array}\right).
\label{eqs6}
\end{equation}
Here $\delta l$ is the size of the spatial grid cell, $\delta t$ is the associated time step, and the top bar stands for an area-average operation. The unit vectors $\mathbf{e}_{p}$ and $\mathbf{e}_{s}$ are defined in such a way so that $\mathbf{e}_{p}=\mathbf{e}_{n}\times\mathbf{e}_{s}$. Evidently, setting $\eta=0$ will recover the standard finite-difference expression where $E_{s}$ and $E_{p}$ are decoupled. Furthermore, $E_{s}, E_{p}, F_{n}, J_{s}, J_{p}$ should sit right at the interface to achieve acceptable numerical accuracy.

\section{Supplementary material II: Discrete Dipole Approximation}

Using the discrete dipole approximation, each metallic cross is approximated as an electric dipole with a dipole polarization $\mathbf{p}=\alpha\mathbf{E}^{l}$. Here $\mathbf{E}^{l}$ is the local electric field, and $\alpha$ is the dipole polarizability. Under a normal incidence $\mathbf{E}^{i}e^{ik_{0}z}$, each cross possesses an identical polarization $\mathbf{p}$. Consequently, the local electric field at the origin can be written as
\begin{equation}
[\underline{\mathbf{I}}+\underline{\mathbf{R}}(k_{0})]\cdot\mathbf{E}^{i}+\sum_{mn\neq00}\underline{\mathbf{G}}^{f}(0,\mathbf{r}_{mn})\cdot\mathbf{p}+\sum_{mn}\underline{\mathbf{G}}^{r}(0,\mathbf{r}_{mn})\cdot\mathbf{p}.
\label{eqs7}
\end{equation}
where $k_{0}$ is the free-space wave number, and $\mathbf{r}_{mn}$ describes the location of the $mn$-th unit cell of the dipole array. It is assumed that the metallic crosses sit right at the $z=0$ plane, and there is no separation between these crosses and the multilayer substrate.

The first term of the local field contains the incident wave plus its reflection by the substrate alone. Because of the magnetoelectric effect, the reflection coefficient $\underline{\mathbf{R}}$ is non-diagonal and is given by
\begin{equation}
\underline{\mathbf{R}}(k_{0z})=\left(\begin{array}{cc}r_{ee}\mathbf{\hat{e}}_{-}\mathbf{\hat{e}}_{+}&r_{eh}\mathbf{\hat{e}}_{-}\mathbf{\hat{h}}_{+}\\ r_{he}\mathbf{\hat{h}}_{-}\mathbf{\hat{e}}_{+}&r_{hh}\mathbf{\hat{h}}_{-}\mathbf{\hat{h}}_{+}\end{array}\right),
\label{eqs8}
\end{equation}
where $k_{0z}=\sqrt{k_{0}^{2}-k^{2}_{x}-k^{2}_{y}}$ with $k_{0z}$ taken such that its imaginary part is positive, $\mathbf{\hat{e}}$ and $\mathbf{\hat{h}}$ are related to the $s$ (transverse electric) and $p$ (transverse magnetic) wave respectively, and $r_{\sigma\sigma'}$ is the reflection coefficient of a process in which an incident $\sigma$ wave is reflected to a $\sigma'$ wave by the substrate \cite{sipe}.

The second term contains the free-space Green's function $\underline{\mathbf{G}}^{f}$, and represents a field due to the dipole-dipole interaction through free space. For a square array of dipoles, it is well known that this electric field can be approximated as $\varsigma\mathbf{p}$ with
\begin{equation}
\varsigma=\frac{z_{0}\omega}{4A_{cell}}\left(\frac{\cos k_{0}R_{0}}{k_{0}R_{0}}-\sin k_{0}R_{0}\right)-iz_{0}\omega\left(\frac{k^{2}_{0}}{6\pi}-\frac{1}{2A_{cell}}\right),
\label{eqs9}
\end{equation}
where $A_{cell}$ is the unit cell area, $R_{0}=\sqrt{A_{cell}}/1.438$, and $z_{0}=\sqrt{\mu_{0}/\epsilon_{0}}$ is the free space impedance \cite{Tretyakov}.

The third term of Eq. (\ref{eqs7}) corresponds an electric field due to the dipole-dipole interaction through the substrate. Using the Green's function $\underline{\mathbf{G}}^{r}(0,\mathbf{r}_{mn})$
\begin{equation}
\underline{\mathbf{G}}^{r}(0,\mathbf{r}_{mn})=\frac{i\omega^{2}\mu_{0}}{8\pi^{2}}\int\int d\mathbf{k}_{s}e^{-i\mathbf{k}_{s}\cdot\mathbf{r}_{mn}}\frac{1}{k_{0z}}\underline{\mathbf{R}}(k_{0z}),
\label{eqs10}
\end{equation}
and the identity
\begin{equation}
\sum_{mn}e^{i\mathbf{k}_{s}\cdot\mathbf{r}_{mn}}=\frac{4\pi^{2}}{A_{cell}}\sum_{mn}\delta(\mathbf{k}_{s}-\mathbf{g}_{mn}),
\label{eqs11}
\end{equation}
where $\mathbf{g}$ being the two-dimensional reciprocal lattice vectors of the dipole array, one can reformulate it as
\begin{equation}
\frac{i\omega^{2}\mu_{0}}{2A_{cell}}\sum_{mn}\frac{1}{k_{0z}}\underline{\mathbf{R}}(k_{0z})\cdot\mathbf{p}\equiv\underline{\mathbf{R}}^{t}\cdot\mathbf{p}.
\label{eqs12}
\end{equation}
By grouping the reciprocal lattice vectors, the above equation can be further simplified as
\begin{equation}
\underline{\mathbf{R}}^{t}\cdot\mathbf{p}=\frac{iz_{0}\omega}{2A_{cell}}\left(\begin{array}{cc}R_{ee}\mathbf{p}&R_{eh}(\mathbf{e}_{z}\times\mathbf{p})\\ R_{he}(\mathbf{e}_{z}\times\mathbf{p})&R_{hh}\mathbf{p}\end{array}\right),
\label{eqs13}
\end{equation}
with
\begin{equation}
R_{ee}=\sum_{m\geq n\geq0}\frac{k_{0}}{k_{0z}}\upsilon r_{ee},\:R_{eh}=\sum_{m\geq n\geq0}\upsilon r_{eh},\:R_{he}=\sum_{m\geq n\geq0}\upsilon r_{he},\:R_{hh}=-\sum_{m\geq n\geq0}\frac{k_{0z}}{k_{0}}\upsilon r_{hh}
\label{eqs14}
\end{equation}
where $\upsilon=1$ when $m=n=0$, $\upsilon=2$ when $m=n$ or $m>n=0$, and $\upsilon=4$ when $m>n>0$. It is important to mention that $k_{0z}$ is purely imaginary for any non-zero reciprocal vector $\mathbf{g}$, because the incident wavelength is bigger than the lattice constant of the dipole array. Consequently, evanescent waves may contribute significantly to $\underline{\mathbf{R}}^{t}$.

Once we know the local electric field, the polarization $\mathbf{p}$ possessed by each metallic cross can be solved as
\begin{equation}
\mathbf{p}=\zeta\left[\underline{\mathbf{I}}-\zeta\underline{\mathbf{R}}^{t}\right]^{-1}\cdot[\underline{\mathbf{I}}+\underline{\mathbf{R}}(k_{0})]\cdot\mathbf{E}^{i}.
\label{eqs15}
\end{equation}
where $\zeta=(\alpha^{-1}-\varsigma)^{-1}$. Using this polarization, one can obtain the total reflected field in the far-field zone
\begin{equation}
\mathbf{E}^{r}(z)=e^{-ik_{0}z}\left\{\underline{\mathbf{R}}(k_{0})\cdot\mathbf{E}^{i}-\frac{i\omega }{2\eta_{0}A_{cell}}\left[\underline{\mathbf{R}}(k_{0})+\underline{\mathbf{I}}\right]\cdot\mathbf{p}\right\}.
\label{eqs16}
\end{equation}


\begin{thebibliography}{99}
\bibitem{code} All necessary codes written in FORTRAN can be made available upon request.
\end{thebibliography}
\end{document}